%% file: ms.tex
\pdfoutput=1
\documentclass{article}

\usepackage{amsmath}
\usepackage{amssymb}
\usepackage{amsthm}
\usepackage{mathtools}
\mathtoolsset{showonlyrefs,showmanualtags}

\usepackage{bbm}
\newcommand{\bbone}{\mathbbm{1}}
\usepackage{graphicx}
\usepackage{microtype}
\usepackage{booktabs}
\usepackage{url}
\usepackage{hyperref}

\usepackage[accepted]{icml2021}

\input{stddef.tex}

\icmltitlerunning{Randomization does not imply unconfoundedness}

\begin{document}

\twocolumn[
\icmltitle{Randomization does not imply unconfoundedness}

\begin{icmlauthorlist}
\icmlauthor{Fredrik Sävje}{yale}
\end{icmlauthorlist}

\icmlaffiliation{yale}{Yale University, New Haven, Connecticut, USA}

\icmlcorrespondingauthor{Fredrik Sävje}{fredrik.savje@yale.edu}

\icmlkeywords{Causal inference, randomization, unconfoundedness}

\vskip 0.3in
]

\printAffiliationsAndNotice{}

\begin{abstract}
A common assumption in causal inference is that random treatment assignment ensures that potential outcomes are independent of treatment, or in one word, unconfoundedness.
This paper highlights that randomization and unconfoundedness are separate properties, and neither implies the other.
A  study with random treatment assignment does not have to be unconfounded, and a study with deterministic assignment can still be unconfounded.
A corollary is that a propensity score is not the same thing as a treatment assignment probability.
These facts should not be taken as arguments against randomization.
The moral of this paper is that randomization is useful only when investigators know or can reconstruct the assignment process.
\end{abstract}

\section{Introduction}

Causal inference is impossible without assumptions, and the predominant assumption to facilitate causal inference is unconfoundedness.
Unconfoundedness goes under other names, such as ignorability and causal exchangeability, but they all describe the setting where treatment is independent of potential outcomes, perhaps after conditioning on covariates.
A common assumption is that randomization ensures that unconfoundedness holds, and randomization is a common way to motivate the unconfoundedness assumption.

This idea is prevalent throughout most subfields of causal inference, as the following quotes illustrate.\footnote{Some of these quotes have been mildly edited for clarity, primarily by substituting mathematical notation with words.}
\citet{Angrist2009Mostly} write ``Random assignment of treatment solves the selection problem because random assignment makes treatment independent of potential outcomes.''
\citet{Wooldridge2010Econometric} writes ``Suppose that the treatment is statistically independent of potential outcomes, as would occur when treatment is randomized across agents.''
\citet{Pearl2018Book} write ``Randomization eliminates confounder bias.''
Finally, \citet{Hernan2020Causal} write ``Randomization is so highly valued because it is expected to produce exchangeability.''

The purpose of this paper is to highlight that randomization does not imply unconfoundedness and that unconfoundedness does not imply that treatment was randomized.
Randomization does allow investigators to estimate treatment effects without systematic errors, but this is true only when they have sufficient knowledge about the assignment mechanism.
The fact that treatment was randomized does not, by itself, ensure that we can estimate treatment effects accurately; knowledge about \emph{how} it was randomized is crucial.
Indeed, provided that the assignment mechanism is randomized and known, inference is possible regardless of whether unconfoundedness holds.
This is useful because randomized assignment mechanisms that are confounded may improve precision over mechanisms that are unconfounded.

Parts of what will be said in this paper is already known by some people working on causal inference.
Presumably, this group includes some of the authors quoted above.
However, practitioners appear to not fully appreciate the fact that randomization and unconfoundedness are distinct concepts.
The confusion likely stems from the fact that some of the most popular assignment mechanisms happen to ensure that unconfoundedness holds, and the term ``randomization'' is often used as shorthand for this subset of mechanisms.
As a consequence, practitioners have been led to believe that any type of randomization will produce unconfoundedness, and they estimate their treatment effects accordingly.
This confusion provides the motivation for the discussion in this paper.

\section{Preliminaries}

A few stylized examples will suffice to make the points of this paper, but the points apply more widely.
Consider an infinite population of units indexed by the unit interval.
Each unit $i$ has two potential outcomes $y_i(1), y_i(0)$, two covariates $x_i, u_i$, and a treatment indicator $w_i$.
All these variables are binary.
The usual assumptions are made on the potential outcomes to ensure that they are well-defined, including no interference and no hidden version of treatment.

The population is such that all units have $y_i(0) = 0$.
The population is then equally divided between the eight possible combination of values of $y_i(1)$, $x_i$ and $u_i$.
For example, $12.5\%$ of the units have $(y_i(1), u_i, x_i) = (1, 0, 1)$, and so on.
Treatment $w_i$ is potentially randomly assigned in the population, and we will discuss its assignment in detail later.

We draw a random sample of $n$ units from the population uniformly and independently.
The sample observations are indexed by the set $\braces{1, \dotsc, n}$.
To differentiate between the units in the population and in the sample, we will denote variables pertaining to sample observations with capital letters.
That is, $Y_i(1)$, $Y_i(0)$, $X_i$, $U_i$ and $W_i$ are the variables for the $i$th sample observation, which has no connection to unit $i$ in the population.
The index $i$ does not carry any information in the sample, so it will be dropped when convenient.
But the index could potentially carry information in the population, so it will never be dropped when referring to population variables.

We do not observe all variables in the sample.
We observe $Y_i$, $X_i$ and $W_i$, where $Y_i = W_i Y_i(1) + (1 - W_i) Y_i(0)$ is the realized outcome.
In other words, one of the potential outcomes and the covariate $U_i$ are unobserved.

We are now ready to formally define the relevant concepts.

\begin{definition}\label{def:randomized}
	A study is \emph{randomized} if $\gamma < \Pr{w_i = 1} < 1 - \gamma$ for all units $i$ in the population and some $\gamma > 0$.
\end{definition}

\begin{definition}
	A \emph{treatment probability} is the probability that an individual unit $i$ in the population is treated $\Pr{w_i = 1}$.
\end{definition}

\begin{definition}\label{def:unconfounded}
	A study is \emph{unconditionally unconfounded} if the potential outcomes are independent of treatment assignment in the sample: $(Y(1), Y(0)) \indep W$. A study is \emph{conditionally unconfounded} if potential outcomes are conditionally independent of treatment assignment given observed covariates: $(Y(1), Y(0)) \indep W \mid X$.
\end{definition}

\begin{definition}\label{def:p-score}
	A \emph{propensity score} is the probability that a sample observation with a certain covariate value is treated $\Pr{W = 1 \given X = x}$.
\end{definition}

Note that probability operators on population variables, such as $\Pr{w_i = 1}$, do not involve sampling variability.
However, probability operators on sample variables, such as $\Pr{W_i = 1}$, involve variability both from treatment assignment in the population and sampling.
Therefore, $\Pr{w_i = 1}$ and $\Pr{W_i = 1}$ will generally not take the same value.

Unconfoundedness is here defined as full independence between both potential outcomes and treatment.
In many cases, this can be weakened to independence of conditional expectation functions.
The arguments in this paper are unaffected by such a weakening.

The current setup has treatment assignment taking place in the population before sampling.
This was chosen to conform with the usual i.i.d.\ setting.
An alternative setup is to first sample units and then assign treatment in the sample.
As long as treatment is assigned independently between units, which is the case for all assignment mechanisms consider here, the difference between assignment in the population or sample is immaterial.
For more intricate assignment mechanisms that induce dependence between units' treatments, the difference can be important, but such an investigation is beyond the scope of this paper.
Nevertheless, a similar argument would still apply.

\section{Confounded random assignment}\label{sec:conf-rand}

Consider an assignment mechanism in the population that assigns treatment independent at random with the following treatment probabilities:
\begin{equation}
	\Pr{w_i = 1} = \frac{3 - y_i(1)}{4}.\label{eq:conf-rand-prob}
\end{equation}
That is, a unit with $y_i(1) = 0$ will have a $75\%$ probability of being treated, and a unit with $y_i(1) = 1$ will have a $50\%$ probability.
Clearly, this satisfies Definition~\ref{def:randomized}, so the study is randomized.

But this study is not unconfounded.
To see this, note that unconfoundedness requires that $\Pr{Y(1) = 1 \given W = 1}$ is the same as $\Pr{Y(1) = 1 \given W = 0}$.
However, by Bayes' theorem, the probability $\Pr{Y(1) = 1 \given W = w}$ is equal to
\begin{equation}
	\frac{\Pr{W = w \given Y(1) = 1} \Pr{Y(1) = 1}}{\Pr{W = w}},
\end{equation}
which in turn is equal to $2 / (3 + 2w)$.
In other words, a treated sample observation has a $40\%$ probability of having $Y(1) = 1$, while an untreated sample observation has a $66.7\%$ probability of having $Y(1) = 1$.
Clearly, this does not satisfy unconditional unconfoundedness.
Conditioning on $X$ does not change matters, and the study remains confounded.

This setting also highlights that a propensity score is different from a treatment probability.
As given by Definition~\ref{def:p-score}, the propensity score is $\Pr{W = 1 \given X = x}$.
Using the law of total probability and the fact that $Y(1)$ and $X$ are independent, we can write the score as
\begin{multline}
	\Pr{Y(1) = 1} \Pr{W = 1 \given Y(1) = 1}
	\\
	+ \Pr{Y(1) = 0} \Pr{W = 1 \given Y(1) = 0},
\end{multline}
which evaluates to $5/8$.
Equation~\eqref{eq:conf-rand-prob} tells us that the treatment probabilities are either $1/2$ or $3/4$, but they are never $5/8$.
Clearly, propensity scores are not the same as treatment probabilities.

Standard estimation techniques will not accurately estimate the treatment effect in this setting.
The average treatment effect $\E{Y(1) - Y(0)}$ is here $1/2$, but the difference-in-means estimator,
\begin{equation}
	\frac{\sum_{i=1}^n W_iY_i}{\sum_{i=1}^n W_i} - \frac{\sum_{i=1}^n (1-W_i)Y_i}{\sum_{i=1}^n (1-W_i)}, \label{eq:dif-est}
\end{equation}
will concentrate around the value $2/5$.
Estimators that make covariate adjustments will not change matters.

\section{Unconfounded deterministic assignment}

Consider an assignment mechanism that assigns treatment so that $w_i = 1$ for all units with $u_i = 1$, and $w_i = 0$ for all units with $u_i = 0$.
That is, we have $w_i = u_i$ for all units in the population.

Definition~\ref{def:randomized} is not satisfied here, so this study is not randomized.
All treatment probabilities are either zero or one, meaning that we would not observe any variability in the treatments assigned to the units in the population upon repeated draws from the assignment mechanism.
This is sometimes referred to as deterministic assignment or lack of positivity.
The consequence is that conventional estimation methods for randomized experiments that use information about treatment probabilities cannot be used in this setting \citep[see, e.g.,][]{Aronow2013Class}.

Still, this study is unconfounded.
Because $Y(1)$ is binary and $Y(0)$ is zero for all units, it is enough to show that $\Pr{Y(1) = 1 \given W = w}$ is constant in $w$ to prove unconditional unconfoundedness.
Because $W = U$ with probability one, we can consider the conditional probability of $Y(1)$ given $U$ instead.
Recall that half of the units with $U = 1$ had $Y(1) = 1$, and half of the units with $U = 0$ also had $Y(1) = 1$.
Hence, irrespective of $w$, we have
\begin{equation}
	\Pr{Y(1) = 1 \given W = w} = 1/2.
\end{equation}
The conclusion is that the study is unconfounded.

Furthermore, while all treatment probabilities are either zero or one, both the conditional and unconditional propensity scores are one half:
\begin{equation}
	\Pr{W = 1 \given X = x} = \Pr{W = 1} = 1/2,
\end{equation}
for all $x$ in the support of $X$.
Hence, the usual overlap condition, which stipulates that the propensity score is bounded away from zero and one, is satisfied.
This shows that we can have overlap without positivity.

What this means is that the difference-in-means estimator in Equation~\eqref{eq:dif-est} is unbiased for the average treatment effect.
The only complication is when we happen to draw a sample that contains only treated units or only control units.
However, the probability of that event diminishes exponentially in the sample size, so it can safely be ignored as long as the sample is not very small.

Some readers might question the practical relevance of this example.
Unconfoundedness rests on a knife edge here: exactly half of the units in each stratum of $y_i(1)$ have $u_i = 1$.
While it is unlikely that naturally occurring populations can sustain balancing acts like this, the purpose of the current example is to illustrate that it is possible, at least in principle, to have a deterministic assignment mechanism that is unconfounded.
In applied research, skepticism towards deterministic assignment mechanisms is sensible.

\section{Unconfounded random assignment}\label{sec:unconf-rand}

There are some randomized assignment mechanisms that do produce unconfoundedness.
One such mechanism is when the treatment probabilities are constant over the units in the population.
That is, $\Pr{w_i = 1} = p$ for all units in the population for some $p \in (0, 1)$.
In this case, the study is unconditionally unconfounded, and the propensity score equals the treatment probability $p$.
Presumably, most of the authors cited in the introduction had an assignment mechanism like this in mind.

A slightly more intricate assignment mechanism sets the unit-level treatment probability according to some function depending on the observable covariates.
That is, we have $\Pr{w_i = 1} = f(x_i)$ for all units in the population for some function $f$ bounded away from zero and one.
In this case, the study is unconfounded conditional on $X$, and the propensity score equals the function deciding treatment probabilities:
\begin{equation}
	\Pr{W = 1 \given X = x} = f(x),
\end{equation}
for all $x$ in the support of $X$.
Of course, it is here important that the function deciding treatment probabilities only depends on observed covariates; the study might not be unconfounded if it depends on the unobserved covariate $u_i$.

While these assignment mechanisms ensure unconfoundedness, additional restrictions are required to ensure that we can estimate the treatment precisely.
In particular, nothing has yet been said about the joint assignment probabilities, but they are critical for the behavior of any estimator.
As an example, consider an assignment mechanism that flips one fair coin and assigns either all units to treatment or all units to control depending on the outcome of the coin flip.
Here, the unit-level treatment probabilities are all $\Pr{w_i = 1} = 1/2$, and the study is unconfounded.
Furthermore, overlap holds because both the unconditional and conditional propensity scores are $1/2$.
The issue is that we never observe a sample containing both treated and control units, because the population only ever contains units from one of the treatment groups.
We might have an unbiased estimator of the treatment effect here, but its precision will not noticeable improve with the sample size.

A simple solution is to require that the assignment mechanism is such that treatment is independently assigned in the population.
Provided that the propensity score is known or can be estimated well, an assignment mechanism like this would ensure that the treatment effect can be estimated with precision in large samples.
This is what \citet{Imbens2015Causal} refer to as a ``regular assignment mechanism,'' but there are many commonly occurring assignment mechanisms that do not fit this mold.

\section{Known assignment mechanism}

The fact that randomization does not imply unconfoundedness may seem to lessen the value of randomization, and indeed, it does.
As the example in Section~\ref{sec:conf-rand} shows, randomization alone is not very helpful.
It is when we have sufficiently knowledge about the assignment mechanism that randomization shines.

To show this, let $p_i = \Pr{w_i = 1}$ denote the treatment probability for unit $i$ in the population, and let $P_i$ be the treatment probability of the $i$th sample observation.
Note that $P_i$ is not the same as the unconditional propensity score $\Pr{W_i = 1}$, nor is it the same as the conditional propensity score $\Pr{W_i = 1 \given X_i}$.
Although, as noted above, they sometimes coincide.

If we observe $P_i$, and it is bounded away from zero and one, we can use the following inverse probability weighting-type estimator to estimate the average treatment effect:
\begin{equation}
	\frac{1}{n} \sum_{i=1}^n \frac{W_iY_i}{P_i} - \frac{1}{n} \sum_{i=1}^n \frac{(1-W_i)Y_i}{1-P_i}. \label{eq:ht-est}
\end{equation}
This is sometimes called the Horvitz--Thompson estimator after \citet{Horvitz1952Generalization}, but it was first described by \citet{Narain1951}.

The power of a randomized assignment mechanism is that it ensures that this estimator is unbiased for the average treatment effect no matter how the treatment probabilities were decided upon.
This is true even if the treatment probabilities are strongly, or even perfectly, correlated with the potential outcomes.
This can be shown by an application of the law of iterated expectation:
\begin{equation}
	\E[\bigg]{\frac{WY}{P}}
	= \E[\bigg]{ \frac{Y(1)}{P} \E[\big]{W \given P, Y(1)} }
	= \E{ Y(1) },
\end{equation}
where we used $\E{W \given P, Y(1)} = P$.
Further restrictions on the assignment mechanism are needed to ensure concentration, but they are fairly mild.

Inverse probability weighting-type estimators can also be used with propensity scores, in which case we would replace $P_i$ in Equation~\eqref{eq:ht-est} with $\Pr{W_i = 1 \given X_i}$, or an estimate thereof.
The two estimators appear similar, but the similarities are deceptive.
Provided that $0 < p_i < 1$ for all units, inverse probability weighting using $P$ will be unbiased for the treatment effect no matter how treatment otherwise was assigned.
If we observe $P$, we do not need to impose any restriction on the assignment mechanism, nor do we need any more information about it.
However, inverse probability weighting using $\Pr{W_i = 1 \given X_i}$ will be successful only when conditional unconfoundedness given $X$ holds.
That is, propensity score weighting requires a specific type of assignment mechanism, and its applicability is therefore more limited.

In a sense, $P$ acts as an universal control variable, always providing the information we need to make credible causal inferences no matter what the assignment mechanism happen to be.
One might be tempted to interpret $P$ as just another covariate and say that the study is unconfounded conditional on it.
However, this perspective misses the critical point that $P$ does not primarily provide information about the units, as ordinary covariates do.
Instead, it provides information directly about the assignment mechanism.

Some readers may find this an interesting curiosity but ask about the practical relevance of the result.
Why would one ever pick an assignment mechanism other than the ones described in Section~\ref{sec:unconf-rand}?
One reason is that investigators may not be in control of treatment assignment.
There could be budgetary, ethical or practical considerations that make an unconfounded mechanism impossible to implement.
There are also studies that leverage naturally occurring assignment mechanisms that are known to be randomized, but are completely beyond the control of the investigator.

Another reason is that assignment mechanisms that do not produce unconfoundedness can potentially perform better than unconfounded mechanisms.
In particular, precision can often be improved by inducing exactly the type of correlation between treatment and potential outcomes that renders the study confounded.
We will consider a slightly different stylized setting to illustrate this.
Such a stylized setting is not necessary to make these points, but it greatly eases the exposition.

Suppose that $y_i(0) = 0$ and $0 < y_i(1) < 2 \E{Y(1)}$ for all units in the population.
We will compare two assignment mechanisms.
The first mechanism sets $p_i$ to be proportional to the potential outcome under treatment.
In particular, it sets $p_i = y_i(1) / 2 \E{Y(1)}$.
The second mechanism sets all treatment probabilities to one half: $p_i = 1/2$.
For both mechanisms, the assignments are independent between units.
Note that both mechanisms ensure that half of the units are treated in expectation, $\E{P} = 1/2$, so any differences cannot be explained by having access to more or less observations of treated units.
The difference is instead that the first mechanism induces a strong correlation between treatment and the potential outcome, rendering it confounded, while the second  mechanism makes the study unconfounded.

The estimator given in Equation~\eqref{eq:ht-est} is unbiased under both assignment mechanisms, so we will focus on its precision.
Because the sample is drawn independently and treatment is assigned independently, the terms of the estimator are independent.
Hence, the normalized variance under both mechanisms is
\begin{equation}
n \Var[\big]{\widehat\tau} = \Var[\bigg]{ \frac{WY}{P} }.
\end{equation}
This is, however, where the similarities end.

For the first mechanism, we have $P = Y(1) / 2 \E{Y(1)}$ with probability one, so
\begin{equation}
	\Var[\bigg]{ \frac{WY}{P} } = 4 \Var{ W } \braces[\big]{\E{Y(1)}}^2.
\end{equation}
We have $\Var{ W } = 1/4$ because $\E{P} = 1/2$, so the normalized variance is $n \Varsub[\big]{1}{\widehat\tau} = \braces[\big]{\E{Y(1)}}^2$ under the first mechanism.

The derivation is somewhat more intricate for the second mechanism.
First, use the law of total variance to write the variance as
\begin{equation}
	\Var[\Big]{ 2 Y(1) \E{ W } } + \E[\Big]{ 4 (Y(1))^2 \Var{ W } }
\end{equation}
Because $P$ is constant at one half here, we have $\E{W} = 1/2$ and $\Var{W} = 1/4$, and the normalized variance under the second mechanism is
\begin{equation}
	n \Varsub[\big]{2}{\widehat\tau} = \Var[\big]{ Y(1) } + \E[\big]{ (Y(1))^2 }.
\end{equation}

The difference in variances for the two assignment mechanisms is therefore
\begin{equation}
	n \Varsub[\big]{2}{\widehat\tau} - n \Varsub[\big]{1}{\widehat\tau} = 2 \Var[\big]{ Y(1) }.
\end{equation}
The variance is always greater under the second mechanism, and the difference can be substantial.
If our aim is to maximize precision, we should pick the confounded assignment mechanism.

The difference is even starker for other estimators or sampling procedures.
All uncertainty under the first mechanism can be ascribed to variability in the number of treated units in the sample, something which the estimator given in Equation~\eqref{eq:ht-est} does not account for.
If the sampling procedure is modified so that exactly half of the sampled units are treated, the variance is zero under the first mechanism.
That is, we learn the treatment effect without error.
The normalized variance is $2 \Var{ Y(1) }$ under the second mechanism, which is smaller than before, but typically larger than zero.
An estimator that adjusts for the number of sampled treated units will have a similar effect.
Examples of such estimators are described by \citet{Hajek1971} and \citet{Imbens2004Nonparametric}.

Of course, it is impossible to implement the first assignment mechanism in practice because it would require perfect knowledge of the potential outcomes.
But this is beside the point.
The point is instead to show that introducing correlation between the potential outcomes and treatment can be enormously beneficial.
While this particular mechanism is not possible to implement, we can hope to emulate it, and thereby possibly reap some of its benefits.

\section{Discussion}

At the heart of this paper is the insight that a study's sampling procedure is different from its treatment assignment mechanism.
Much of the causal inference literature treats them as one and the same.
This is a consequence of the practice of making i.i.d.\ assumptions by default.

The perspective that the actual characteristics of the sampling procedure and assignment mechanism should be taken into account in analysis is sometimes referred to as ``design-based.''
The design-based perspective has a long history in survey sampling, as described by for example \citet{Saerndal1992Model}.
The perspective is increasingly popular in causal inference \citep[see, e.g.,][]{Freedman2008Regression,Aronow2013Class,Lin2013Agnostic,Imbens2015Causal}.
One aim of this paper is to illustrate the usefulness, and often necessity, of this perspective.

The design-based perspective can be interpreted as an extension of the frequentist approach to statistical inference.
Probability statements are here not only seen as statements about limits of relative frequencies in an infinite sequence of trials, but investigators are also expected to explicitly specify what device generated those trials.
An abstract stream of observations from an unknown source will not cut it.
Therefore, investigators must ask from where the stochasticity in their studies comes, and analyze the data accordingly.
Doing so will avoid mistakes, including assuming that randomization always provides unconfoundedness.
Making the data generating process concrete tends to also make inferences clearer and more relevant.

Points related to the ones made here have recently been made by \citet{Abadie2020Sampling} and by \citet{Titiunik2021Natural}.
The first set of authors also highlights the difference between sampling and treatment assignment mechanism, but focuses on the precision of regression estimators when variability from treatment assignment is explicitly taken into account.
The discussion by \citet{Titiunik2021Natural} is largely parallel to the one in this paper, highlighting that random assignment does not imply unconfoundedness in the context of natural experiments.\footnote{I thank Peter Aronow for making me aware of the chapter by \citet{Titiunik2021Natural} shortly before the start of the workshop.}
These discussions complement the current paper, and interested readers will find them valuable.

\section*{Acknowledgements}

I thank Peter Aronow, Josh Kalla, Winston Lin and Jas Sekhon for helpful comments.

\bibliographystyle{icml2021}
\bibliography{references}

\end{document}

%% file: stddef.tex
\DeclarePairedDelimiter\braces\lbrace\rbrace

\providecommand{\bbone}{\mathbf{1}}
\DeclarePairedDelimiterXPP\indicator[1]{\bbone}{\lbrack}{\rbrack}{}{#1}

\DeclarePairedDelimiterXPP\expf[1]{\exp}{\lparen}{\rparen}{}{#1}
\DeclarePairedDelimiterXPP\logf[1]{\log}{\lparen}{\rparen}{}{#1}
\DeclarePairedDelimiterXPP\maxf[1]{\max}{\lparen}{\rparen}{}{#1}
\DeclarePairedDelimiterXPP\minf[1]{\min}{\lparen}{\rparen}{}{#1}

\DeclarePairedDelimiterXPP\func[2]{#1}{\lparen}{\rparen}{}{#2}

\makeatletter
\newcommand*{\tran}{{\mathpalette\@tran{}}}
\newcommand*{\@tran}[2]{\raisebox{\depth}{$\m@th#1\intercal$}}
\makeatother

\DeclarePairedDelimiterXPP\tnorm[1]{}{\lVert}{\rVert_{1}}{}{#1}
\DeclarePairedDelimiterXPP\enorm[1]{}{\lVert}{\rVert_{2}}{}{#1}
\DeclarePairedDelimiterXPP\inorm[1]{}{\lVert}{\rVert_{\infty}}{}{#1}
\DeclarePairedDelimiterXPP\pnorm[2]{}{\lVert}{\rVert_{#1}}{}{#2}

\DeclarePairedDelimiterXPP\detf[1]{\det}{\lparen}{\rparen}{}{#1}

\DeclareMathOperator{\trsym}{tr}
\DeclarePairedDelimiterXPP\tr[1]{\trsym}{\lparen}{\rparen}{}{#1}

\DeclareMathOperator{\diagsym}{diag}
\DeclarePairedDelimiterXPP\diag[1]{\diagsym}{\lparen}{\rparen}{}{#1}

\DeclareMathOperator{\ranksym}{rank}
\DeclarePairedDelimiterXPP\rank[1]{\ranksym}{\lparen}{\rparen}{}{#1}

\DeclareMathOperator{\vectorizesym}{vec}
\DeclarePairedDelimiterXPP\vectorize[1]{\vectorizesym}{\lparen}{\rparen}{}{#1}

\providecommand\given{}
\newcommand\givensymbol[1]{\nonscript\:#1\vert\allowbreak\nonscript\:\mathopen{}}

\let\Prsym\Pr
\let\Pr\relax
\DeclarePairedDelimiterXPP\Pr[1]{\Prsym}{\lparen}{\rparen}{}{%
	\renewcommand\given{\givensymbol{\delimsize}}%
	#1}

\DeclarePairedDelimiterXPP\Prsub[2]{\Prsym_{#1}}{\lparen}{\rparen}{}{%
	\renewcommand\given{\givensymbol{\delimsize}}%
	#2}

\DeclareMathOperator{\Esym}{E}
\DeclarePairedDelimiterXPP\E[1]{\Esym}{\lbrack}{\rbrack}{}{%
	\renewcommand\given{\givensymbol{\delimsize}}%
	#1}

\DeclarePairedDelimiterXPP\Esub[2]{\Esym_{#1}}{\lbrack}{\rbrack}{}{%
	\renewcommand\given{\givensymbol{\delimsize}}%
	#2}

\DeclareMathOperator{\Varsym}{Var}
\DeclarePairedDelimiterXPP\Var[1]{\Varsym}{\lparen}{\rparen}{}{%
	\renewcommand\given{\givensymbol{\delimsize}}%
	#1}

\DeclarePairedDelimiterXPP\Varsub[2]{\Varsym_{#1}}{\lparen}{\rparen}{}{%
	\renewcommand\given{\givensymbol{\delimsize}}%
	#2}

\DeclarePairedDelimiterXPP\EstVar[1]{\widehat{\Varsym}}{\lparen}{\rparen}{}{%
	\renewcommand\given{\givensymbol{\delimsize}}%
	#1}

\DeclareMathOperator{\Covsym}{Cov}
\DeclarePairedDelimiterXPP\Cov[1]{\Covsym}{\lparen}{\rparen}{}{%
	\renewcommand\given{\givensymbol{\delimsize}}%
	#1}

\DeclarePairedDelimiterXPP\Covsub[2]{\Covsym_{#1}}{\lparen}{\rparen}{}{%
	\renewcommand\given{\givensymbol{\delimsize}}%
	#2}

\DeclareMathOperator{\Corrsym}{Corr}
\DeclarePairedDelimiterXPP\Corr[1]{\Corrsym}{\lparen}{\rparen}{}{%
	\renewcommand\given{\givensymbol{\delimsize}}%
	#1}

\makeatletter
\newcommand{\indep}{\protect\mathpalette{\protect\@indep}{\perp}}
\newcommand*{\@indep}[2]{\mathrel{\rlap{$#1#2$}\mkern3mu{#1#2}}}
\makeatother

\newcommand{\bigOsym}{\mathcal{O}}
\DeclarePairedDelimiterXPP\bigO[1]{\bigOsym}{\lparen}{\rparen}{}{#1}

\newcommand{\littleOsym}{o}
\DeclarePairedDelimiterXPP\littleO[1]{\littleOsym}{\lparen}{\rparen}{}{#1}

\newcommand{\bigOpsym}{\bigOsym_p}
\DeclarePairedDelimiterXPP\bigOp[1]{\bigOpsym}{\lparen}{\rparen}{}{#1}

\newcommand{\littleOpsym}{\littleOsym_p}
\DeclarePairedDelimiterXPP\littleOp[1]{\littleOpsym}{\lparen}{\rparen}{}{#1}

\newcommand{\bigOmegasym}{\Omega}
\DeclarePairedDelimiterXPP\bigOmega[1]{\bigOmegasym}{\lparen}{\rparen}{}{#1}

\newcommand{\littleOmegasym}{\omega}
\DeclarePairedDelimiterXPP\littleOmega[1]{\littleOmegasym}{\lparen}{\rparen}{}{#1}

\newcommand{\bigThetasym}{\Theta}
\DeclarePairedDelimiterXPP\bigTheta[1]{\bigThetasym}{\lparen}{\rparen}{}{#1}

\theoremstyle{plain}

\theoremstyle{definition}

\newtheorem{definition}{Definition}

\theoremstyle{remark}